\documentclass[aps,pre,groupedaddress,showpacs,preprint]{revtex4}
\usepackage{graphicx}
\usepackage[dvips]{epsfig}
\usepackage{dcolumn}
\usepackage{bm}
\usepackage{amssymb}
\usepackage{amsmath}

\begin{document}
\title{Outbreaks of Hantavirus induced by seasonality}
\author{C. Escudero$^{\dag}$, J. Buceta$^{\ddag}$, F. J. de la
Rubia$^{\dag}$, and Katja Lindenberg$^{\ddag}$ }

\affiliation{
$^{\dag}$Departamento de F\'{\i}sica Fundamental,
Universidad Nacional de Educaci\'on a Distancia, C/ Senda del Rey 9, 28040 Madrid, Spain\\
$^{\ddag}$Department of Chemistry and Biochemistry, and Institute
for Nonlinear Science, University of California San Diego, 9500 Gilman Dr., La Jolla, CA 92093-0340, USA}

\begin{abstract}
Using a model for rodent population dynamics, we study 
outbreaks of Hantavirus infection induced by the alternation of seasons. 
Neither season by itself satisfies the environmental requirements for
propagation of the disease. This result can be explained in terms of
the seasonal interruption of the relaxation process 
of the mouse population toward equilibrium, and may shed light on the
reported connection between climate variations and outbreaks of the disease.

\end{abstract}

\maketitle

\section{Introduction}
\label{sec1}

Hantaviruses are rodent-borne zoonotic agents that may
cause diseases in humans such as hemorrhagic fever with renal syndrome and 
hantavirus pulmonary syndrome~\cite{mills1,hjelle,hooper}. Hantaviruses have
been identified at an increasing rate in recent years, and as of now about
thirty different ones have been discovered throughout the world. 
One of these, the Sin Nombre virus, was not isolated until 1993
after an outbreak in the Four Corners Region in the USA~\cite{mills1,hjelle}.
The host of this particularly dangerous virus is the deer mouse,
{\it Peromyscus maniculatus}, the most numerous mammal in North America.
The virus produces a chronic infection in the mouse population, but it is not
lethal to them. It is believed that transmission in the rodent population is
horizontal and due to fights, and that the subsequent infection
of humans, where the mortality rate can be as high as fifty percent, is
produced by their contact with the excreta of infected mice. Moreover, so
far there is no vaccine or effective drug to prevent or treat the
hantavirus pulmonary syndrome. Therefore, a major effort has been made
to understand the population dynamics of deer mice colonies in order
to design effective prevention policies~\cite{mills1}.

It has been noted that environmental conditions are directly connected to
outbreaks of Hanta~\cite{hjelle}. For instance, the 1993 and 1998
outbreaks occurring in the Four Corners Region have been associated with the
so-called $El$ $Ni\tilde{n}o$ Southern Oscillation~\cite{hjelle}. Related to
this and other such observations, the effects of seasonality in
ecological systems have been a subject of recent
interest~\cite{selgrade,king}. Multi-year oscillations of mammal
populations~\cite{schaffer}, prey-predator seasonal dynamics~\cite{lima},
and persistence of parasites in plants between seasons~\cite{gubbins}
are examples that illustrate the importance of seasonality in
population dynamics.

Recently, Abramson and Kenkre have proposed a phenomenological model for
mice population that successfully reproduces some features of
Hantavirus propagation~\cite{abramson}.
In particular, that model explores the relation
between resources in the medium, carrying capacity, and the spread of the
infection in the rodent colony. Herein we study the effects of
seasonality in that
model. Our motivation is not only to provide
more realism to the model, but also to investigate the counterintuitive
effects that dynamic alternation may cause in a biological system. Brownian
motors~\cite{astumian} and switching-induced morphogenesis~\cite{buceta} are
examples that show that alternation in time of ``uninteresting'' dynamics may
produce ``interesting'' outcomes. Along these lines, we will show that
alternation of
seasons, neither of which by itself fulfills the environmental requirements
on the
carrying capacity for spreading of the infection, may produce an
outbreak of the disease. The
mechanism driving this behavior is the interruption of the relaxation
process that equilibrates the mouse population from season to season: if the
duration of seasons becomes shorter than the relaxation time of the
population, the disease spreads.

The paper is organized as follows. In Sec.~\ref{sec2} we briefly review
the model
for mouse populations introduced in~\cite{abramson}. In Sec.~\ref{sec3}
we explain
how seasonality is introduced in that model and analyze the conditions
for Hanta outbreaks to take place due to the alternation of seasons.
The exact solution
of the model and a particular example that illustrates the phenomenology
is given in Sec.~\ref{sec4}. 
Finally, in Sec.\ref{sec5}, we summarize the main conclusions and
propose some directions for future work.

\section{The Model}
\label{sec2}

The model introduced in~\cite{abramson} for the temporal evolution of a 
population of mice subjected to the Hantavirus infection reads: 
\begin{subequations}
\begin{eqnarray}
\frac{dM_{S}}{dt} &=&bM-cM_{S}-\frac{M_{S}M}{K}-aM_{S}M_{I},  \label{model1}
\\
\frac{dM_{I}}{dt} &=&-cM_{I}-\frac{M_{I}M}{K}+aM_{S}M_{I},  \label{model2}
\end{eqnarray}
\end{subequations}
where $M_{S}$ and $M_{I}$ are the population densities of susceptible and
infected mice respectively, $M=M_{S}+M_{I}$ is the total population
of mice, and $a$, $b$, $c$ and $K$ are positive constants. The terms on the
right hand sides of Eqs.~(\ref{model1}) and (\ref{model2}) take into account
the following processes: births with rate constant $b$, depletion by
death with rate constant
$c$, competition for the resources in the medium characterized by the carrying
capacity $K$, and transmission of the infection with rate coefficient $a$.
It is worth
noting that infected pregnant mice produce Hanta antibodies that keep
their fetus free from the infection; that is, {\it all} mice are born
susceptible~\cite{mills1}, as indicated by the absence of a birth term in
Eq.~(\ref{model2}).
Note also the absence of a recovery term in the model since, as mentioned
earlier, mice become chronically infected by the virus.

\begin{figure}[tbp]
\psfig{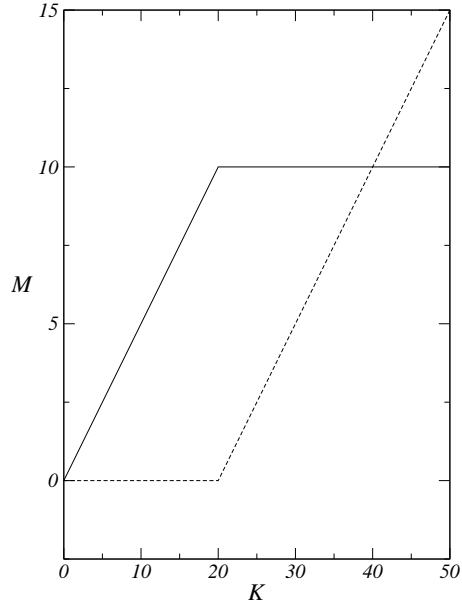}
\caption{ Stable equilibrium population of susceptible (solid line) and
infected (dashed line) mice as a function of the resources present in the
medium, $K$.
The values of the parameters are the same as those used in
Ref.~\cite{abramson}: $a$=0.1,
$b$=1, and $c$=0.5. The value of the critical carrying capacity is $K_{c}=20$.
\label{Fig1}}
\end{figure}

The system of equations (\ref{model1}) and (\ref{model2}) has four
equilibrium points. Two of them are irrelevant for the analysis: the null state
$M_{I}=M_{S}=0$, which is always unstable if $b>c$ (a condition that we will
assume throughout this paper), and a meaningless state with $M_{I}<0$ for any
value of the parameters. The other two equilibria are 
\begin{equation}
M_{S}=K(b-c)
\quad
\quad
M_{I}=0,  \label{equilibrio1}
\end{equation}
\begin{equation}
M_{S}=\frac{b}{a}
\quad
\quad
M_{I}=K(b-c)-\frac{b}{a},  \label{equilibrio2}
\end{equation}
The stability of the equilibrium points (\ref{equilibrio1}) and (\ref{equilibrio2}) depends on the
value of the carrying capacity. If $K<K_{c}=\frac{1}{a}\left( \frac{b}{b-c}\right)$
then (\ref{equilibrio1}) is stable and (\ref{equilibrio2})
unstable. If $K>K_{c}$ it is the other way around. That is,
when the available resources, $K$, are below the critical value, $K_{c}$, the infection
does not propagate in the colony, the whole population of mice grows
healthy, and its size increases proportionally with those resources. As
soon as
$K$ surpasses $K_{c}$ the
virus spreads in the colony, the susceptible mouse population saturates, and
the fraction of infected mice becomes larger as $K$ increases (see Fig.~\ref
{Fig1}).

\section{Seasonal Alternation}
\label{sec3}

The Four Corners Region, where an important number of cases of Hantavirus
Pulmonary Syndrome have occurred, has a desert climate. The largest climate
variations within this region come from the alternation between dry and
rainy seasons. We will therefore assume alternation in time of these two
seasons. It is important to remark that a two-season assumption
is not crucial, and that the analysis with four seasons is
also straightforward within the 
formalism introduced herein. During each of the two seasons under
consideration we assume there to be no climate variations, so that each
season can be
characterized by a set of time-independent parameters
$\{\rho_{i}\}=\{a_i,b_i,c_i,1/K_i\}$ where
$i=1,2$. We implement square-periodic season alternation where
the duration of each season is $T/2$. Again, this particular alternation
pattern is not
essential for the mechanism. Other schemes of season
alternation, e.g. different duration of the seasons or random switching
between seasons, do not qualitatively change the phenomenology.

Any quantity $\rho(t)$ alternating in the way described above can be written as: 
\begin{equation}  \label{alternation}
\rho(t)=\rho_++\rho_-\mu(t),
\end{equation}
where $\rho_\pm=\frac{1}{2}(\rho_1\pm\rho_2)$ and $\mu(t)$ is a periodic
square wave 
\begin{equation*}
\mu(t)=\left\{ 
\begin{array}{r@{\quad:\quad}l}
1 & 0<t<\frac{T}{2} \\ 
-1 & \frac{T}{2}<t<T
\end{array}
\right..
\end{equation*}
Let us now suppose the following conditions for the sets $\{\rho_{i}\}$
according to seasonality:
\begin{equation}  \label{conditions}
a_1<a_2,\quad b_1<b_2, \quad c_1<c_2, \quad K_1>K_2,
\end{equation}
where 1 stands for the rainy season and 2 for the dry one. The biological
motivation for these conditions is the following. The dry season provides
less resources for the colony than the rainy season $\left(K_{2}<K_{1}\right)$, and as a consequence
the death rate is higher $\left(c_{2}>c_{1}\right)$, and the
transmission rate is also larger since
fights for the available resources are expected to
increase $\left(a_{2}>a_{1}\right)$. However, notice that we consider the birth
rate to be larger during the dry season $\left(b_{2}>b_{1}\right)$.
There are two reasons for this.
First is the assumption that the colony makes an attempt to maintain its
population. Second is an
implementation of the maturation process in the model: baby mice
do not contribute to the propagation of the disease~\cite{mills1},
and, therefore, even if births were actually more
numerous during the rainy season, the contribution of this fact to the
propagation of the infection will only be important in the next
(dry) season, when mice have matured and are ready to fight.
It will be shown later that these assumptions lead
to a situation with high population during the rainy season and low
population during the dry one, in agreement with the available data~\cite{mills1,hjelle}.

Moreover, we assume that $K_1<\min(K_{c1},K_{c2})$, i.e., the 
resources are {\it all times} (during both seasons) {\it below} the minimum
critical threshold that triggers the propagation of the disease. We will
show that nevertheless it is possible for the infection to spread.

\begin{figure}[tbp]
\psfig{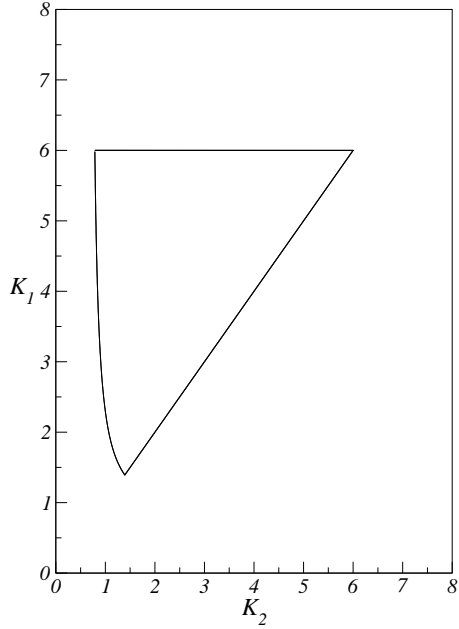}
\caption{$(K_1,K_2)$-region of disease propagation due to seasonal
alternation. The
values of the other relevant parameters are given
in Eqs.~(\ref{parameters1}) and (\ref{parameters2}). Every value of
$K_1$ or $K_2$  within
the enclosed area by itself does {\em not}
satisfy the required environmental conditions that support spreading of
the virus in the mice colony.
Yet, those same points lead to outbreaks of Hanta when alternation
of seasons is sufficiently fast.
\label{Fig2}}
\end{figure}

The equilibrium populations of the susceptible and infected mice are determined by the
set of values $\{\rho_{i}\}$. When switching from season to season,
the populations evolve
trying to reach a new equilibrium. Therefore, the dynamics is driven by
the competition
between two characteristic times. On the one hand there
is an {\em external} time scale
determined by the seasonal forcing, $t_{e}=T/2$. On the other hand, the
relaxation toward
equilibrium after a switching of seasons involves a relaxation time.
The latter measures
the time required for the mouse colony to relax to the equilibrium
state associated with $\{\rho_{j}\}$
after having been driven during the previous season by the
conditions $\{\rho_{i}\}$, that is,
$t_{r}(i\rightarrow j)$ where $i,j=1,2$. The {\em internal} time
scale is defined
as the fastest relaxation process, i.e., $t_{i}=\min [t_{r}(i\rightarrow j)]$.
``Fast'' or ``slow'' seasonal alternation then refers to the
comparison between these two time scales. If $t_{e}\gg t_{i}$
the mouse population has enough time to accommodate to the new
conditions from season
to season and relax to equilibrium. Moreover, since we have imposed
the condition that the resources
at any time of the year are below the critical thresholds $K_{ci}$,
there will be no infected mice. In the other limit, $t_{e}\ll t_{i}$, seasonal
changes occur too fast, the relaxation process is
interrupted, and no equilibrium can be reached from season to season.
Then note that an adiabatic elimination can be
implemented~\cite{gardiner}, and
$\mu (t)$ in Eq.~(\ref{alternation}) can be replaced
by its average value, $\left\langle \mu (t)\right\rangle =0$. Therefore, in
the limit of fast season alternation the system is driven by the
set of averaged values $\{\rho_{+}\}=\{a_{+},b_{+},c_{+},(1/K)_{+}\}$,
and the critical
carrying capacity is given by $K_{c+}=\frac{b_{+}}{a_{+}(b_{+}-c_{+})}$.
As a consequence, it is possible to find regions of parameters
where $K_{c+}$ is smaller than the {\it effective} value of the carrying
capacity associated with the averaged values: 
\begin{equation*}
\left[ \left( \frac{1}{K}\right) _{+}\right] ^{-1}=\left[\frac{1}{2}\left(\frac{1}{K_{1}}+\frac{1}{K_{2}}\right)\right]^{-1}=\frac{2K_{1}K_{2}}{K_{1}+K_{2}},
\end{equation*}
and the infection propagates.

General conditions leading to this behavior can be posed, but the
expressions are rather cumbersome.
We prefer, for the sake of simplicity, to show a particular typical case.
We use the following values for the parameters:
\begin{equation}
a_{1}=\frac{1}{4}, \quad b_{1}=1, \quad c_{1}=\frac{1}{3},  \label{parameters1}
\end{equation}
\begin{equation}
a_{2}=4, \quad b_{2}=\frac{73}{72}, \quad c_{2}=1,  \label{parameters2}
\end{equation}
that lead to $K_{c1}=6$ and $K_{c2}=73/4$ respectively.
The dynamics are completely determined once the value of the
carrying capacity during each season is specified.
According to the previous discussion,
these parameters can be chosen such that the following conditions hold:
\begin{equation*}
\left[ \left(1/K\right) _{+}\right] ^{-1}>K_{c+}, \quad K_{1}>K_{2},
\quad K_1<\min(K_{c1},K_{c2}).
\end{equation*}
These conditions lead to the points $\left(K_{1},K_{2}\right)$
that fulfill the seasonal requirements
given by Eq.~(\ref{conditions}), so that slow alternation of seasons leads
to infection-free states while fast alternation leads to Hanta outbreaks.
This region is plotted in Fig.~\ref{Fig2}. Notice in particular that
the point $\left(K_{1}=4,K_{2}=1\right)$
lies inside the region and that $K_{i}<K_{ci}$. In the next section
we illustrate the seasonality-induced propagation of the disease
for this particular point.

\section{The Critical Period}
\label{sec4}

So far we have determined that outbreaks of Hanta induced entirely by
seasonal changes can occur if the duration of the seasons are short
enough. Now we establish the meaning of ``short enough'' quantitatively.
Since $K_{c+}$ is strictly smaller than the effective value of the carrying
capacity, there should be a finite value of $T_c$ such that for any $T<T_{c}$
the population of infected mice is greater than zero, but for periods
above this critical period the infected population goes to zero.

In order to obtain the value of the critical period we need to solve the
system of equations (\ref{model1}) and (\ref{model2}). In spite of its
nonlinearities the system can be solved analytically by means of a
reciprocal transformation \cite{kenkre} and the following exact
solution is obtained: 
\begin{subequations}
\begin{eqnarray}
M_{I}(t,I_{0},S_{0};\{\rho \})&=&\frac{I_{0}\Omega (t)}{S^{aK-1}+aI_{0}\int_{0}^{t}\Omega (\tau )d\tau },   \label{solution1} \\
M_{S}(t,I_{0},S_{0};\{\rho \})&=&\frac{SM_{0}e^{\frac{S}{K}t}}{(\Omega
(t)e^{ct})^{\frac{1}{aK-1}}}-\frac{I_{0}\Omega (t)}{S^{aK-1}+aI_{0}\int_{0}^{t}\Omega (\tau )d\tau },  \label{solution2}
\end{eqnarray}
\end{subequations}
where $I_{0}$ and $S_{0}$ are the initial conditions for $M_{I}$ and $M_{S}$
respectively, and the following definitions have been introduced,
\begin{equation*}
\Omega (t)=e^{-ct}\left( M_{0}\left( e^{\frac{S}{K}t}-1\right) +S\right)
^{aK-1},
\end{equation*}
\begin{equation*}
S=K(b-c), \qquad M_{0}=I_{0}+S_{0}.
\end{equation*}
Because the external forcing due to the alternation of seasons is
periodic, Floquet theory ensures the existence of a periodic solution.
The values of $I_{0}$ and $S_{0}$ compatible with the non-equilibrium
periodic solution can
be obtained by evolving the system during the first half of a period under
dynamics $1$ and the second half under
dynamics $2$, and forcing periodity on the solutions after a
whole period of evolution, that is,
\begin{subequations}
\begin{eqnarray}
M_{I}\left(\frac{T}{2},M_{I}\left(\frac{T}{2},I_{0},S_{0};\{\rho _{1}\}\right),M_{S}\left(\frac{T}{2},I_{0},S_{0};\{\rho _{1}\}\right);\{\rho _{2}\}\right)&=&I_{0}, \label{algebra1} \\  
M_{S}\left(\frac{T}{2},M_{I}\left(\frac{T}{2},I_{0},S_{0};\{\rho _{1}\}\right),M_{S}\left(\frac{T}{2},I_{0},S_{0};\{\rho _{1}\}\right);\{\rho _{2}\}\right)&=&S_{0}.  \label{algebra2}
\end{eqnarray}
\end{subequations}
In order to close the system in the non-equilibrium steady state
$M_{I}\left(t,T;\{\rho _{1,2}\}\right)$ and $M_{S}\left(t,T;\{\rho _{1,2}\}\right)$,
the values of $I_{0}$ and $S_{0}$ that solve that system of equations~(\ref{algebra1}) and (\ref{algebra2}) must be
then re-introduced in Eqs.~(\ref{solution1}) and (\ref{solution2}).
The critical period is then the largest value of $T$ satisfying the condition 
\begin{equation*}
M_{I}\left(t,T;\{\rho_{1,2}\}\right)>0.
\end{equation*}

\begin{figure}[tbp]
\psfig{file=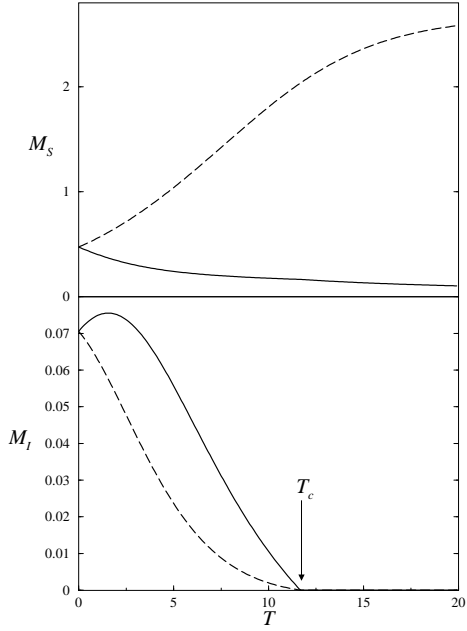,width=6cm,angle=0}
\caption{ Population of susceptible (top) and infected (bottom) mice versus
the period of the seasons. The dashed and continuous lines indicate the populations
at the end of the rainy and dry seasons respectively.
The critical period for which the virus begins to spread due to seasonality is $T_c \simeq 12$.
The values of the relevant parameters are
$a_{1}=1/4$, $b_{1}=1,c_{1}=1/3$, $K_{1}=4$
and $a_{2}=4$, $b_{2}=73/72$, $c_{2}=1$, $K_{2}=1$ for the rainy
and dry seasons respectively. \label{Fig3}}
\end{figure}

\begin{figure}[tbp]
\psfig{file=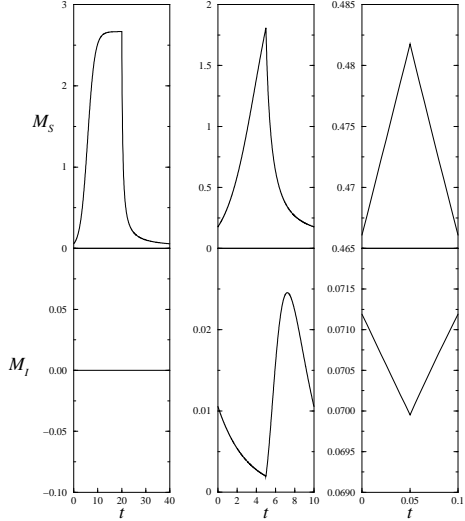,width=6cm,angle=0}
\caption{Population of susceptible (top) and infected (bottom) mice versus
time for a period of evolution. The values of the relevant parameters
are the same as those
used in Fig.~\ref{Fig3}. The critical period is $T_c \simeq 12$. Left panel: Results for a very long
period ($T=40$). Right panel: Very short period ($T=0.1$). Central panel : Results for a near-critical period ($T=10$). \label{Fig4}}
\end{figure}

We illustrate the procedure with the example mentioned above where
the parameters are given by Eqs.~(\ref{parameters1}) and (\ref{parameters2}),
and with $K_1=4$ and $K_2=1$.
The results are shown in Figs.~\ref{Fig3} and \ref{Fig4}.
The values of $M_I$ and $M_S$ as a function of the period of the seasons
are depicted in Fig.~\ref{Fig3}, where the populations of the susceptible
and infected mice at the end of each season are given. As seen
in that figure, the value of the critical period is $T_c \simeq 12$.
Note that if the alternation is slow, $T>T_c$, all mice grow healthy.
On the other hand, if the alternation is faster than the relaxation time
required by the colony to accommodate its population from season to season, $T<T_c$, the
virus spreads and $M_{I}>0$. Notice that in the limit $T\rightarrow 0$
the dynamics is driven by
$\{\rho_{+}\}$ and the populations of susceptible and infected mice
are given by Eq.~(\ref{equilibrio2})
with $a=a_{+}$, $b=b_{+}$, $c=c_{+}$, and $1/K=(1/K)_{+}$. Let us stress again that the carrying
capacity is below its critical threshold at any time. 

In Fig.~\ref{Fig4} we plot, for different period lengths,
the exact solutions $M_{I}\left(t,T;\{\rho _{1,2}\}\right)$ and $M_{S}\left(t,T;\{\rho _{1,2}\}\right)$
as a function of time through one period of evolution.
The first semi-period corresponds to the rainy season and the second to the dry season.
When seasons last long (left panel), there are no infected mice and
the susceptible population
simply oscillates between the two equilibrium states given by Eq.~(\ref{equilibrio1}).
For sufficiently short seasons (right panel), there is propagation of
the disease and the values of $M_I$ and $M_S$ fluctuate around the
equilibrium points
determined by Eq.~(\ref{equilibrio2}) and the set of parameters $\{\rho_{+}\}$.
Finally, when the period of the seasons is near, but below, the critical
period (central panel), the
infected population is small and the population $M_{I}$ oscillates
in a more pronounced manner.

\section{Conclusions}
\label{sec5}

By introducing seasonality in a paradigmatic model for Hantavirus
propagation in mice colonies, we have shown that the alternation of seasons
may cause outbreaks of the disease. The striking feature of that behavior
lays in the fact that neither season satisfies the conditions for the
infection to spread in terms of the availability of resources. 
The mechanism responsible for the phenomenon is the
competition between two time scales: an external one, the duration of a
season, and an internal one, the relaxation time for the mouse colony to
equilibrate its population from season to season. We have shown that
if the duration of the seasons is longer than the relaxation time,
no propagation of Hantavirus occurs. On the other hand, if the
relaxation process is
interrupted by a fast seasonal alternation, the disease spreads. 
We have analyzed the general conditions for which the phenomenon occurrs.
Moreover, we have illustrated the mechanism with a particular example
that can be solved exactly.

This work may help to clarify the reported relation between climate and
propagation of Hanta in mice populations. However, to elucidate whether the
proposed phenomenon actually takes place in nature we depend on data that
unfortunately are not available in the literature. One can envision
further modifications of
the model that may improve its features, such as, for
example, the inclusion of spatial dependence or of noisy contributions to the
dynamics. Finally, we stress that the general idea underlying
the mechanism is model-insensitive and can therefore be extended to other
systems where seasonality plays a relevant role.
Work along these directions is in progress.

\section*{Acknowledgments}

The authors gratefully acknowledge fruitful comments from and discussions
with V. M. Kenkre during the elaboration of this work.
C. Escudero is grateful to the Department of Chemistry and Biochemistry of
the University of California, San Diego for its hospitality. This work has
been partially supported by
the National Science Foundation under grant PHY-9970699, by
the Ministerio de Educaci\'{o}n y Cultura (Spain)
through grants No. AP2001-2598 and EX2001-02880680,
and by the Ministerio de Ciencia y Tecnolog\'{i}a (Spain),
Project No. BFM2001-0291.

\end{document}